# Zero-field spin-splitting and spin lifetime in *n*-InSb/In$_{1-x}$Al$_x$Sb asymmetric quantum well heterostructures

A. M. Gilbertson[1,2], M. Fearn[1], J. H. Jefferson[1], B. N. Murdin[3], P. D. Buckle[1], L. F. Cohen[2]

[1]*QinetiQ, St. Andrews Road, Malvern, Worcestershire, WR14 3PS, United Kingdom*
[2]*Blackett Laboratory, Imperial College London, Prince Consort Rd., London, SW7 2BZ, United Kingdom*
[3]*Advanced Technology Institute, University of Surrey, Guildford, GU2 7XH, United Kingdom*

The spin-orbit (SO) coupling parameters for the lowest conduction subband due to structural (SIA) and bulk (BIA) inversion asymmetry are calculated for a range of carrier densities in [001]-grown δ-doped *n*-type InSb/In$_{1-x}$Al$_x$Sb quantum wells using the established 8 band ***k · p*** formalism [PRB **59**,8 R5312 (1999)]. We present calculations for conditions of zero bias at 10 K. It is shown that both the SIA and BIA parameters scale approximately linearly with carrier density, and exhibit a marked dependence on well width when alloy composition is adjusted to allow maximum upper barrier height for a given well width. In contrast to other material systems the BIA contribution to spin splitting is found to be of significant and comparable value to the SIA mechanism in these structures. We calculate the spin lifetime $\tau_{s[1\bar{1}0]}$ for spins oriented along $[1\bar{1}0]$ based on D'yakonov-Perel' mechanism using both the theory of Averkiev *et al.* [J. Phys.:Condens. Matter **14** (2002)] and also directly the rate of precession of spins about the effective magnetic field, taking into account all three SO couplings, showing good agreement. $\tau_{s[1\bar{1}0]}$ is largest in the narrowest wells over the range of moderate carrier densities considered, which is attributed to the reduced magnitude of the *k*-cubic BIA parameter in narrow wells. The inherently large BIA induced SO coupling in these systems is shown to have considerable effect on $\tau_{s[1\bar{1}0]}$, which exhibits significant reduction in the maximum spin lifetime compared to previous studies which consider systems with relatively weak BIA induced SO coupling. The relaxation rate of spins oriented in the [001] direction is found to be dominated by the k-linear SIA and BIA coupling parameters and at least an order of magnitude greater than in the $[1\bar{1}0]$ direction.

## I. INTRODUCTION

Spin-orbit (SO) induced spin splitting of the conduction bands of III-V semiconductors has been studied extensively both theoretically and experimentally [1-8], with more recent efforts linked with the growing interest in the field of spintronics [9,10] and the development of a spin-based field effect transistor [11]. Narrow bandgap semiconductors such as InSb offer interesting and novel characteristics for these applications such as small effective mass, inherently large SO coupling and large effective Landé *g*-factor. Of particular relevance to spin transistor applications, the small effective mass gives rise to high electron mobility, and hence fast, low power devices [12]. The large SO coupling mixed across a small gap results in a large Rashba effect (see below) which means that a large modulation of the spin lifetime with electric field is possible. Even though the large SO coupling results in a small spin lifetime in 2D structures, reducing the spin-diffusion length, this is mitigated by the small effective mass.

It is now understood that the SO coupling in asymmetric quantum well (QW) heterostructures results from the lack of inversion symmetry that originates from two predominant sources, which lift the spin degeneracy even in the absence of an applied magnetic field. The first is an intrinsic property of the zinc-blende crystal structure called bulk inversion asymmetry (BIA) as described by Dresselhaus [13]. In bulk semiconductors, this mechanism leads to a spin splitting proportional to the wave vector cubed, ***k***$^3$, characterized by the coefficient γ. However, in a spatially inhomogeneous structure such as a QW where the *z* axis is in the growth direction, $k_z$ is replaced by $-i\partial/\partial z$ within the envelope function approximation (EFA) and the BIA term is separated into two contributions; a ***k***-cubic term and a ***k***-linear term characterized by the coefficient β [1].

The second source of inversion asymmetry is due to a change in the macroscopic potential and bandstructure through a heterojunction interface, known as structural inversion asymmetry (SIA) as described by Bychkov and Rashba [14]. This produces an additional spin-splitting linear in ***k*** and characterized by the coefficient α. The combined SO Hamiltonian for SIA and BIA induced couplings for a two dimensional electron gas (2DEG) confined in the *x-y* plane is of the form [1,15]

$$H_{SO}(\boldsymbol{k}_\parallel) = \alpha(k_x\sigma_y - k_y\sigma_x) + \beta(k_x\sigma_x - k_y\sigma_y) + \gamma(k_x^2 k_y\sigma_y - k_x k_y^2\sigma_x), \qquad (1)$$

where $\sigma_x$ and $\sigma_y$ are the Pauli spin matrices. The spin splitting induced by the SO coupling in Eq. 1 can described by introducing a $k$-dependent *effective* magnetic field $\boldsymbol{B}_{eff}(\boldsymbol{k}_\parallel)$ about which spins precess which an effective Larmor frequency $\boldsymbol{\Omega}(\boldsymbol{k}_\parallel) = g^*\mu_B \boldsymbol{B}_{eff}(\boldsymbol{k}_\parallel)/\hbar$, where $\mu_B$ is the Bohr magnetron and $g^*$ is the effective Landé $g$-factor. The corresponding Hamiltonian describing the SO coupling can now be rewritten in the form $H_{SO}(\boldsymbol{k}_\parallel) = \hbar\boldsymbol{\sigma}\cdot\boldsymbol{\Omega}(\boldsymbol{k}_\parallel)$ analogous to a Zeeman term, where $\boldsymbol{\sigma}$ is the vector of Pauli spin matrices and $\boldsymbol{\Omega}(\boldsymbol{k}_\parallel)$ is given by

$$\boldsymbol{\Omega}(\boldsymbol{k}_\parallel) = \frac{1}{\hbar}[(\beta k_x - \alpha k_y - \gamma k_x k_y^2), (\alpha k_x - \beta k_y + \gamma k_x^2 k_y), 0]. \qquad (2)$$

The rate of spin precession is thus determined by the magnitude of the SO interactions through the coupling parameters $\alpha$, $\beta$ and $\gamma$. For this reason the SIA mechanism, or 'Rashba effect', has been subject to considerable attention for device applications as the heterostructure potential and hence the magnitude of $\alpha$ can be influenced by an electric field due to an external gate electrode. This effect was first exploited in the spin field effect transistor (S-FET) proposal by Datta and Das [11]. As pointed out by D'yakonov and Perel' [16] the decoherence of an initially aligned (polarized) population of spins is driven by their individual precssion about the vector $\boldsymbol{\Omega}(\boldsymbol{k}_\parallel)$. For a spin polarized current in a diffusive system, the presence of momentum scattering randomly changes the wave vector $\boldsymbol{k}_\parallel$ in the time $\tau_p$ which in turn changes the orientation of the effective magnetic field, and precession vector. For frequent momentum scattering events the average precession frequency and hence spin relaxation rate is slowed (motional narrowing) leading to a characteristic spin relaxation rate $\tau_s \propto \tau_p^{-1}$. This is the D'yakonov-Perel' (DP) spin relaxation mechanism [16], shown to be present in most III-V semiconductor systems over a broad range of temperatures [17,18,19,20]. It is essential for spintronic devices as the dominance of this mechanism is necessary for the gate modulation of spin populations [21]. Recent experimental results appearing to be consistent with the DP mechanism, indicate that it is dominant for high mobility $n$-type InSb/InAlSb QWs over a broad range of temperatures [22]. Experimental evidence for the gate control of $\alpha$ has now been demonstrated by the beating of Shubnikov-de Haas (SdeH) oscillations of magnetoresistance [6,7], quantum interference effects such as weak anti-localisation in low field magnetoresistance [23], and more recently, the modification of spin lifetime with gate has been attributed to SIA [21,24]. Although many of these measurements are indirect and would benefit from a robust model which incorporates realistic device structures as described here.

Zero-field Rashba splitting has previously been studied in symmetric InSb QWs in the high barrier limit by Stanley *et al.* [25] using Kane's 8-band $\boldsymbol{k}\cdot\boldsymbol{p}$ theory within the EFA. It was found that the spin-splitting was up to a factor of two greater than that of equivalent InAs QWs due to the larger split-off and smaller band gap energies. Here we extend the work of Ref. 25 to more realistic $n$-InSb/In$_{1-x}$Al$_x$Sb asymmetric heterostructures with finite barriers, considering realistic growth constraints such as critical thickness strain relaxation, providing a comprehensive study of the influence of heterostructure design on the SO coupling parameters in these structures. Critically, we take into account the penetration of the wave function into the barriers, shown to add a dominant contribution to the Rashba coefficient $\alpha$ [3,14], and which is shown here to also influence $\beta$. Previously, it has been assumed that SIA is the dominant source of zero field spin-splitting in narrow gap semiconductors [5,8,15]. In this article we present theoretical results counter to this assertion, demonstrating that both the $k$-linear and $k$-cubic BIA terms can give a significant contribution toward the total spin splitting and should not be neglected.

In addition to this we calculate the DP spin lifetimes for these heterostructures, considering the modifications to the spin lifetime field effect transistor (SL-FET) as proposed by Cartoxia *et al.* [26] and Schliemann *et al.* [27] in this material system with realistic coupling parameters. These devices are based on the properties of the spin lifetime tensor in the DP regime due to the interference between BIA and SIA interactions. Following the theoretical results of Averkiev *et al.* [28], the novel SL-FET exploits the tunable Rashba effect to achieve the conditions of $\alpha = \beta$ and $\alpha \neq \beta$, in order to significantly change the spin lifetime and provide current modulation. This change occurs since for $\alpha = \pm\beta$ the direction of $\boldsymbol{\Omega}(\boldsymbol{k}_\parallel)$ (with $\gamma = 0$) no longer depends on $k$ i.e. $\boldsymbol{\Omega}(\boldsymbol{k}_\parallel) \propto \alpha(\pm k_x - k_y)[1, \pm 1, 0]$ (from Eq. 2) so that scattering events change only its magnitude and not direction, allowing long spin lifetimes for spins oriented in the [110] and [1$\bar{1}$0] directions respectively, i.e. when parallel to $\boldsymbol{\Omega}(\boldsymbol{k}_\parallel)$ [28]. Although the SL-FET concept has received some criticism as being a counter productive incarnation of the original Datta-Das S-FET [29], it is interesting to investigate the viability of balancing the SIA and BIA interactions for the purpose of enhancing the spin lifetime in the InSb/ In$_{1-x}$Al$_x$Sb QW system.

This paper is organized in the following way. In Sec. II we present the 8-band **k · p** model of Pfeffer et al. [5,15] used for the calculations of coupling coefficients. Sec. III contains the results of the calculated Rashba and Dresselhaus coefficients where the effects of QW width $W$, doping density $N_d$, spacer thickness $S$ (see Fig. 1) and barrier composition on the magnitude of $α$, $β$ and $γ$ are investigated. In Sec. IV the spin-splitting is calculated and the relative contributions from SIA and BIA is discussed. In Sec. V we calculate the DP spin lifetimes [30] in these heterostructures using the obtained SO coupling parameters in the regimes of constant and varying carrier density where the effect of the inherently large SO coupling in this 2D system is manifested in a significant reduction in the maximum spin lifetime compared to previous studies in wider gap III-V compounds [28,31]. Finally in Sec. VI the paper is concluded with a summary of the results.

## II. THE **k · p** MODEL

To calculate the conduction band SO coupling parameters, we employ the 8-band **k · p** model described by Pfeffer et al. [15] for its successful modelling of experimental results of III-IV semiconductors such as the extensively studied GaAs/GaAlAs heterostructures [8] and recently achieved agreement with spin resonance results obtained from InSb/In$_{0.91}$Al$_{0.09}$Sb asymmetric QWs by Khodaparast et al. [5,32], of particular relevance to the heterostructures studied here.

The three-level model **k · p** Hamiltonian for the conduction, valence (light and heavy hole) and split-off bands respectively, results in an 8 x 8 differential matrix (including spin) which is reduced by Gaussian elimination to a 2 x 2 differential matrix for the conduction band. This model has been shown to accurately describe the conduction bands of narrow gap semiconductors such as InSb, as the coupling between the conduction and valance/split-off bands dominates [5,33]. The resulting Schrödinger-like equation for the [001] growth direction is [5]

$$\begin{pmatrix} \hat{A}-\lambda & \hat{K} \\ \hat{K}^* & \hat{A}-\lambda \end{pmatrix} \begin{pmatrix} \Psi_1(z) \\ \Psi_2(z) \end{pmatrix} = 0, \quad (3)$$

where $\lambda$ is the energy eigenvalue and

$$\hat{A} = -\frac{\hbar^2}{2}\frac{\partial}{\partial z}\frac{1}{m^*(z,\lambda)}\frac{\partial}{\partial z} + \frac{\hbar^2 k_\perp^2}{2m^*(z,\lambda)} + V(z). \quad (4)$$

The off-diagonal term consists of two parts $\hat{K} = \hat{K}_{SIA} + \hat{K}_{BIA}$ for the SIA and BIA interactions given by

$$\hat{K}_{SIA} = -i\sqrt{2}k_-\hat{\alpha}(z,\lambda), \quad (5)$$

$$\hat{K}_{BIA} = -i\sqrt{2}k_x k_y k_-\hat{\gamma}(z,\lambda) + \sqrt{2}k_+\hat{\beta}(z,\lambda). \quad (6)$$

where,

$$\hat{\alpha}(z,\lambda) = \frac{P_0^2}{3}\frac{\partial}{\partial z}\left(\frac{1}{\tilde{\varepsilon}_i(z,\lambda)} - \frac{1}{\tilde{f}_i(z,\lambda)}\right), \quad (7)$$

$$\hat{\beta}(z,\lambda) = -\frac{\partial}{\partial z}\tilde{\gamma}(z,\lambda)\frac{\partial}{\partial z}, \quad (8)$$

$$\hat{\gamma}(z,\lambda) = \frac{2BP_0}{3}\left(\frac{1}{\tilde{\varepsilon}_i(z,\lambda)} - \frac{1}{\tilde{f}_i(z,\lambda)}\right), \quad (9)$$

with $\tilde{\varepsilon}_i(z,\lambda) = -E_{gi} + V(z) - \lambda$, $\tilde{f}_i(z,\lambda) = -E_{gi} - \Delta_i + V(z) - \lambda$ and $k_\pm = (k_x \pm i k_y)/\sqrt{2}$. $E_{gi}$ and $\Delta_i$ are the band gap and split-off energies at the zone centre of the materials where the subscript $i$ refers to the different regions of the heterostructure i.e. upper barrier (UB), QW and lower barrier (LB). $\tilde{\varepsilon}_i(z,\lambda)$ and $\tilde{f}_i(z,\lambda)$ thus describe the valence band edge and split-off band edge energy profiles measured relative to $\lambda$. $V(z)$ is the conduction band edge energy profile throughout the heterostructure given by $V(z) = -e\phi(z) + dV_{UB}\Theta(-z) + dV_{LB}\Theta(z+W)$ where $e$ is the electron charge, $\phi(z)$ is the smoothly varying electrostatic potential and $dV_{UB}\Theta(-z)$ and $dV_{LB}\Theta(z+W)$ are step functions characterising the conduction band offsets $dV_{UB}$ and $dV_{LB}$ at the two interfaces located at $z = 0$ and $z = W$ respectively. $B$ is the Kane parameter [34] and $P_0$ is the interband momentum matrix element related to the $\mathbf{k} \cdot \mathbf{p}$ interaction term $E_p$ by $P_0^2 = \hbar^2 E_p / 2m_0$. $B$ and $E_p$ are taken to be -31.4 eVÅ² and 23.1eV ($P_0$ = 9.37eVÅ) respectively for InSb [35] and assumed to be the same for In$_{1-x}$Al$_x$Sb. This is a reasonable approximation since $E_p$ is approximately constant for most III-V compounds [34]. To be consistent with the model of Pfeffer *et al.* [15], potential energies are measured relative to the conduction band edge at $z = 0$ in the QW such that $V_{QW}(0) = 0$ (see Fig. 1).

### III. CALCULATION OF COUPLING PARAMETERS

SO coupling parameters $\alpha$, $\beta$ and $\gamma$ are calculated explicitly by evaluating the expectation values of Eq. 7-9 with respect to $\Psi(z)$ i.e. $\alpha = <\Psi(z)|\hat{\alpha}(z,\lambda)|\Psi(z)>$. The precise form of the SIA and BIA SO Hamiltonians in Eq. 1 has varied subtly between authors by a factor of minus one (see Cartoxia *et al.* [26], Averkiev *et al.*[28] and *Kainz et al.* [31]) which can be absorbed within the coupling parameters. The polarity of the coupling parameters has no effect on the spin-splitting. However, it is important for the direction of $\mathbf{\Omega}(\mathbf{k}_\parallel)$ and the spin lifetime calculations [30]. Here we define the coupling parameters by mapping the off-diagonal matrix elements of Eq. 3 to the form of Eq. 1. Electrostatic potentials $\phi(z)$ and normalised electron wave functions $\Psi(z)$ for each structure are obtained self consistently at 10 K without spin splitting [15] from solutions of $\hat{A}\Psi(z) = \lambda\Psi(z)$, using a 1-band Schrödinger-Poisson model (SPM) tailored to narrow band-gap materials [36]. A typical solution for a 20nm QW is shown in Fig. 1. This SPM is a good approximation since the spin-splitting is small giving $\Psi_1(z) = \Psi_2(z) = \Psi(z)$ [15].

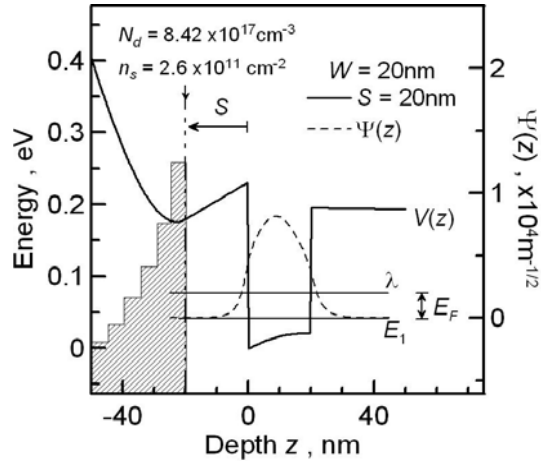

FIG. 1. Conduction band profiles $V(z)$ of a typical δ-doped In$_{0.8}$Al$_{0.2}$Sb/InSb/ In$_{0.85}$Al$_{0.15}$Sb 20nm QW with spacer thickness $S$ = 20nm (left axis), and corresponding normalized wave function $\Psi(z)$ of the 1$^{st}$ subband (right axis) at 10 K. The hatched region represents a schematic of the doping profile used. The doping density $N_d$ and resulting carrier density $n_s$ for this solution are given. The axis is normalized to $V_{QW}(0) = 0$.

It has been shown in real structures to be beneficial for carrier confinement, mobility and surface gate leakage to have alloy composition approaching the maximum barrier height allowed for a given well width [12]. For that reason, the various well widths considered here are modelled with barriers of varying $In_{1-x}Al_xSb$ compositions to best meet this criterion, whilst also maintaining appropriate critical thicknesses for strain relaxation in the QW. Accordingly, Al content in the barrier increases as the well width is narrowed the effect of which is that, counter intuitively, $\alpha$ is larger in wide well structures for a given $n_s$ as discussed in detail below.

TABLE. 1. Parameters used in the SPM calculations at 10 K [37].

| QW width $W$, nm | Upper : lower barrier Al fraction $x$ | Upper : lower barrier energy gaps, meV | Upper : lower barrier effective masses, $m_0$ | Strained InSb QW energy gaps, meV |
|---|---|---|---|---|
| 15 | 0.20 : 0.15 | 651.9 : 548.9 | 0.0342 : 0.0292 | 263.15 |
| 20 | 0.20 : 0.15 | 651.9 : 548.9 | 0.0342 : 0.0292 | 263.15 |
| 25 | 0.18 : 0.13 | 610.7 : 507.7 | 0.0322 : 0.0271 | 260.05 |
| 30 | 0.15 : 0.10 | 548.9 : 445.9 | 0.0292 : 0.0241 | 255.40 |

Table 1 lists the parameters used in the SPM calculations for each QW width considered. For each well width, the doping density $N_d$ and spacer thickness $S$ are varied providing sixteen different SPM solutions for a given $W$. Material parameters used in the SPM calculations were derived from transmission spectroscopy performed by Dai *et al*. [37] to determine the bandgap energies of $In_{1-x}Al_xSb$. The InSb QW is assumed to be lattice matched to the lower barrier and strained accordingly (see Table. 1). A 5% difference in Al composition is maintained in each structure for comparison, and is realistic in ensuring that the top barrier thickness remains below the critical thickness for strain relaxation. A single δ-doping layer is located in the top barrier and simulated by an exponentially decaying dopant profile in the growth direction as a result of secondary-ion-mass-spectroscopy (SIMS) data in these systems, indicated by the shaded region in Fig. 1 [36,38]. The exact functional form of the decay is not expected to affect the results presented here, although its presence is relevant for the positioning of the doping layer above the QW since segregation of dopant atoms into the QW, which can occur when the doping plane is located below the QW, is detrimental to carrier mobility. Notably, within the SPM it is assumed that all donor atoms are ionised and that no inter-diffusion of atoms takes place at the interfaces. $\Delta_{QW}$ is taken as 0.810eV [35] for all QW widths whereas split-off energies for the barriers are estimated from a linear interpolation between the known values for InSb and AlSb (0.750eV [35]) due to the lack of data on the ternary alloys. The conduction : valance band offset ratio is taken as 62% : 38% following Ref. 39. It is assumed in this analysis that the Fermi energy is pinned at mid-gap on the semiconductor surface, consistent with the results of measurements undertaken in similar InSb QW systems [36,39,40]. For all calculations performed here, we restrict ourselves to solutions for zero bias, and to varying the doping parameters $N_d$ and $S$ such that only the lowest subband is occupied.

### A. SIA coefficients

In varying the parameters $N_d$ and $S$ in the calculations we note the acute influence on the conduction band edge potential profile and corresponding wave function asymmetry, defined as the difference between the electron probability density at the interfaces $\Delta\Psi_I^2 = \Psi^2(0) - \Psi^2(W)$. The carrier density in the well $n_s$ is determined by the amount of carriers transferred into the 2DEG which increases with $N_d$ and decreasing $S$ as more dopant is positioned closer to the QW. Therefore we can combine the effects of $N_d$ and $S$ on $\alpha$ by analysis of the variation of $\alpha$ with $n_s$ as presented in Fig. 2(a) for the various well widths. Each data point thus represents a separate SPM solution for varying $N_d$, $S$ and $W$.

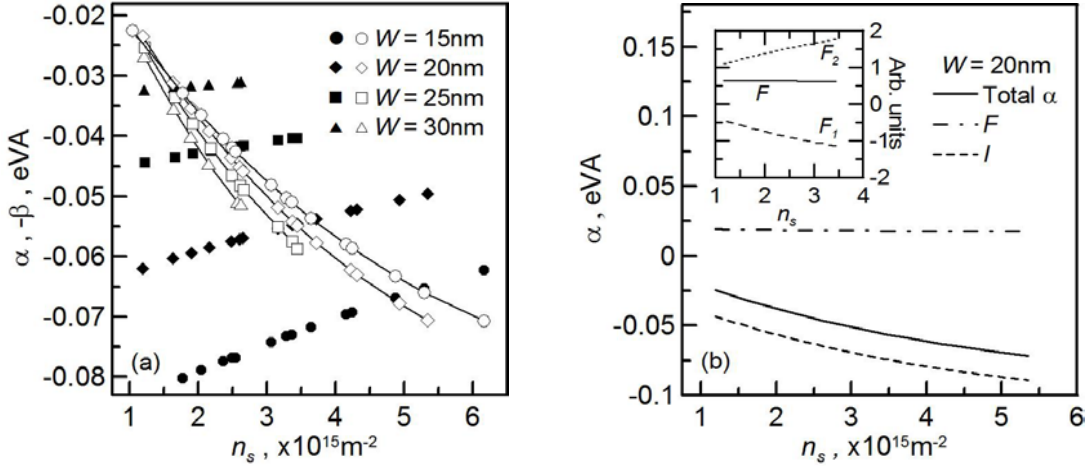

FIG. 2. Calculated results for the SIA SO coupling parameter $\alpha$ as a function of $n_s$ for (a) the various well widths considered (open symbols) along with the k-linear BIA term for comparison (closed symbols), and (b) for a 20nm QW showing the contributions from the 'field' and 'interface' terms. Inset of 2(a) shows the two components of the field term $F$ which result in the observed dependence on $n_s$.

It is found that for small carrier densities the magnitude of $\alpha$ increases approximately linearly with $n_s$ due to an increased conduction band bending in the top barrier. This leads to an increase in the potential asymmetry of the QW, weighting the wave function further towards the top interface and increasing $\Delta\Psi_I^2$ (not presented here). The disparity between $\alpha$ values increases with increasing carrier density in the well. Following Pfeffer et al. [15] the SIA coupling parameter is comprised of a term proportional to the electric field $d\phi(z)/dz$ and interface terms resulting from the derivative of the discontinuous changes in band-edge, $dV_{UB}\Theta(-z)$ and $dV_{LB}\Theta(z+W)$ within $V(z)$. These three terms are given by [15]

$$\alpha = \frac{P_0^2}{3}\left[\left\langle\Psi\left|\frac{\partial\phi}{\partial z}D_i\right|\Psi\right\rangle\right] + \frac{P_0^2}{3}\left[dV_{UB}D_0\Psi^2(0) - dV_{LB}D_W\Psi^2(W)\right] \qquad (10)$$

$$+ \frac{P_0^2}{3}\left[\Psi^2(0)(C_0 - dV_{UB}D_0) - \Psi^2(W)(C_a - dV_{LB}D_W)\right]$$

where $D_i = 1/\tilde{\varepsilon}_i^2 - 1/\tilde{f}_i^2$, $C_0 = \Delta_{QW}/\tilde{\varepsilon}_{QW}(0)\tilde{f}_{QW}(0) - \Delta_{UB}/\tilde{\varepsilon}_{UB}(0)\tilde{f}_{UB}(0)$ and $C_a = \Delta_{QW}/\tilde{\varepsilon}_{QW}(W)\tilde{f}_{QW}(W) - \Delta_{LB}/\tilde{\varepsilon}_{LB}(W)\tilde{f}_{LB}(W)$. The electric field in the conduction band is equal to $Ee = 1/e* d\phi(z)/dz + dV_{UB}\delta(z) - dV_{LB}\delta(z-W)$. Eq. 10 has been transformed in order to match this form by adding and subtracting the term $F_2 = dV_{UB}D_0\Psi^2(0) - dV_{LB}D_W\Psi^2(W)$ (resulting from $<\Psi(z)|EeD_i|\Psi(z)>$) from the right hand side where $D_0$ and $D_W$ are the values of $D_i$ at each interface resulting in the final expression [15]. The first term in Eq. 10 is assigned the label $F_1$ such that the total 'field' term is $F = F_1 + F_2$ (see Fig. 2(b)). The third term in Eq. 10 is the 'interface' term $I$. In agreement with Pfeffer et al. [14] we find that the interface term dominates over the electric field term as shown in Fig. 2(b) for the 20nm QW. The coefficients within the interface term, $C_0 - dV_{UB}D_0$ and $C_W - dV_{LB}D_W$, are negative and vary weakly with $n_s$ compared to $\Psi(0)$ and $\Psi(W)$. As such, we find that the *total contribution to $\alpha$ is proportional to the asymmetry of the electron probability density at the interfaces* ($\Delta\Psi_I^2$). The influence of the electric field can be interpreted as facilitating the asymmetry of the QW profile necessary to weight the wave function toward or away from an interface, thus varying $\Delta\Psi_I^2$. The resulting field term is positive as the inclusion of the electric field at the interfaces in $<\Psi(z)|EeD_i|\Psi(z)>$ results in the addition of a large positive step (the same step is subtracted from the interface term). The observed small variation of $F$ with $n_s$ can be attributed to the competition of the two component field terms $F_1$ and $F_2$ which have opposite dependences on $n_s$ as shown in the inset of Fig. 2(b). It is found that the rate of change in $F_1$ and $F_2$ with $n_s$ varies with well width resulting in a field term $F$ whose dependence on $n_s$ is observed to change as the well width is increased from 15nm. Since $F_2 > F_1$ we find that the electric field in the conduction band is largely determined by the band offsets rather than the average electric field.

The most striking result of Fig. 2(a) is that contrary to previous studies [4,41], for a given carrier density, $\alpha$ is greatest in the wide QW structures, albeit with the drawback of a greatly reduce capacity of the 1st subband compared to the narrow wells, which limits the range of allowed $n_s$. For this reason, larger Rashba coefficients can be achieved in narrower wells at increased carrier densities (see Fig. 2(a)). The increased confinement in narrow wells compared to wider wells had been assumed to increase the wave function penetration into the barriers and hence $\alpha$ (if the underlying asymmetry is the same in case). Whilst this is true for wells of fixed barrier height, in the structures considered here which are a better reflection of real devices, Al content in the barriers and hence barrier heights (with respect to the subband energy $E_1$) are usually increased for the narrower well structures, and the penetration of the wave function into the barriers ($\Delta\Psi_I^2$), and therefore $\alpha$, is reduced compared to the wider wells. The results of Fig. 2(a) also show that the wider wells exhibit an increased sensitivity of $\alpha$ to doping and $n_s$.

These results can be summarised by the assertion that for a single-sided δ-doping scheme, for a given QW width and barrier composition, $\alpha$ is proportional to $n_s$ (where $n_s$ varies as a result of doping). This is true also for δ-doping located in the bottom barrier. However, a distinction between the two is evident when using an external gate to vary $n_s$ and $\alpha$. For structures doped below the QW, the direction of the electric field across the QW is reversed and although $n_s$ increases with positive bias, the QW asymmetry, and hence $\alpha$, decreases as observed in the experimental results of Schapers *et al.* [6] and Nitta *et al.* [7] for InGaAs QWs. Structures doped above the QW exhibit the opposite behaviour of $\alpha$ increasing with positive bias, although this has not been demonstrated experimentally in the literature.

*Comparison with previous experiment and theory*

Our calculated values of $\alpha$ can be compared to previous experimental and theoretical studies of narrow gap InSb and InAs QWs available in the literature. Consistent with previous studies of InSb [5], $\alpha$ is found to be negative. This can also be deduced from inspection of Eq. 10 and the direction of the electric field across the well, whereby, providing that the coefficients $C_0 - dV_{UB}D_0$ and $C_W - dV_{LB}D_W$ are negative, $\alpha$ is negative so long as $\Delta\Psi_I^2 = \Psi^2(0) - \Psi^2(W)$ is positive. It follows from this argument that $\alpha > 0$ for structures doped below the QW as was demonstrated by Nitta *et al.* [7].

The values of $\alpha$ for the well widths considered show good quantitative agreement with the simple one-band model approximation of de Andrada e Silva *et al.* [4] at low carrier densities where the non-parabolicity corrections are small. Khodaparest *et al.* [32] studied spin resonance in symmetric InSb/In$_{0.91}$Al$_{0.09}$Sb QWs where, assuming that all the spin-splitting originated from SIA, estimated a zero field value of $\alpha \sim 0.13$ eVÅ for a well width of 30nm. These results were described by Pfeffer *et al.* [5] using the 8-band $\mathbf{k} \cdot \mathbf{p}$ model described in Sec. II. This value is somewhat larger than our estimates for a similar structure. Grundler [42] investigated the gate dependence of $\alpha$ in symmetric 4nm InAs/InGaAs QWs via the beating of SdeH oscillations extracting values in the range 0.2-0.4eVÅ. These values were estimated using an expression derived by Engels *et al.* [43] relating $\alpha$ to the difference in the populations of the spin-split subbands $\Delta n = n^+ - n^-$ which are determined from Fourier analysis of the beating patterns at low B fields (B < 3T). This approach is somewhat unreliable as it is derived from the modified 2D density of states in the presence of Rashba splitting at B = 0T. Since the zero-field value of $\alpha$ is calculated from transport data in non-zero B fields, the measurement includes contributions from the Zeeman term, hence overestimating $\alpha$. In fact, the Rashba coefficient is rarely measured in narrow gap semiconductors by this method due to the presence of large effective Landé *g*-factors and Zeeman splitting, which dominates the low field magnetoresistance as observed by Refs. [32,44,45]. All of the above methods are based on the assumption that the zero-field spin-splitting is dominated by the SIA mechanism. As is discussed in the following Sec. III, this cannot always be considered true and thus can lead to over estimations of $\alpha$. More recently, a novel experimental approach for the direct measurement of the Rashba and Dresselhaus spin splittings has been proposed by Ganichev *et al.* [46] from the angular dependence of the spin-galvanic photocurrent [47]. Values for the ratio $\alpha / \beta$ in 15nm InAs/InGaAs QWs are estimated as ~2.15 agreeing with ratios from $\mathbf{k} \cdot \mathbf{p}$ calculations in the InGaAs QW system of ~ 1.85 [15]. Ratios calculated here for equivalent QWs are dependent on well width and carrier density but are typically smaller than those quoted above due to the comparatively large Dresselhaus coefficient γ which enhances $\beta$ (see below).

**B. BIA Coefficients**

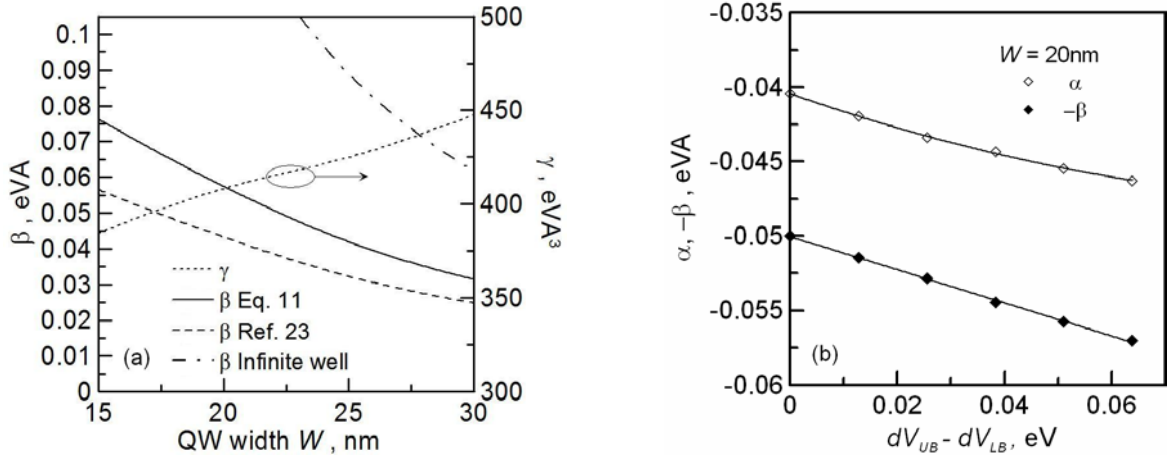

FIG. 3. (a) Calculated results for the BIA SO coupling parameters $\beta$ and $\gamma$ as a function of QW width $W$ for fixed Fermi energy $E_F = 40$meV, showing $\beta$ calculated from $\gamma <k_z^2>$ (dashed line) and $\gamma_{bulk}(\pi/W)^2$ (dot-dashed line) for comparison. (b) Calculated parameters $\alpha$ and $-\beta$ as a function of barrier asymmetry for a 20nm QW with fixed carrier density $n_s = 2.6 \times 10^{11}$cm$^{-2}$ - the solid lines are a guide to the eye.

The BIA parameters $\beta$ and $\gamma$ exhibit a relatively weak linear dependence on $n_s$ and $\lambda$ compared to $\alpha$ for a given well width (demonstrated in Fig. 2(a) for $-\beta(n_s)$) consistent with the results of Kainz et al. [31] for the AlAs/GaAs system. The energy eigenvalue $\lambda$ contains both the confinement energy $E_1$ and the Fermi energy $E_F$ (shown in Fig. 1) thus the dependence on $n_s$ reflects higher order non-parabolicity corrections with increasing Fermi energy [31]. To compare the results for different QW widths, $\beta$ and $\gamma$ are plotted as a function of $W$ in Fig. 3(a) for a *fixed Fermi energy*. The parameter $\gamma$ (dotted line Fig. 3(a)) is observed to increase with well width. For increasing well width, $E_1$ is lowered which results in an enhancement of $\gamma$ and the observed trend. The value of $\gamma$ for the widest well shows a ~30% reduction from that of the bulk, which is estimated as $\gamma_{bulk} = 646.5$eVÅ$^3$ at the zone centre by evaluating Eq. 9 directly with an unstrained energy gap $E_g = 235.2$meV [35] and assuming $\lambda = V(z) = 0$ in the bulk. Reduction from this value is due to the presence of confinement ($E_1$) and an increased density of states in the bulk 3D system which lowers the Fermi energy to within a few kT of the band edge (taken here as zero). Hence for $W \to \infty, E_1 \to 0$, the value of $\gamma$ would approach that of the bulk case. It should be mentioned that our bulk estimation of $\gamma$ is somewhat greater than that estimated from the 16 band **k · p** model of Cardona et al. [48] given as $\gamma^*_{bulk} = 563.9$eVÅ$^3$. This variation can be attributed to the difference in the number of bands accounted for in the models and the uncertainty in the values for the parameter $B$ in the expression for $\gamma$ (Eq. 9). The parameter $B$ is determined either from perturbation theory [49] or experimental measurements of the BIA spin-splitting in the bulk material [50]. The later is somewhat unreliable, predominantly due to the inherent large effective $g^*$-factor in bulk InSb (~-52). As a result, the value of $B$ is shown to vary significantly over a range $-31.4 < B < -12.6$ (eVÅ$^2$) [36,49,50]. The value of $B$ taken from Landolt and Bornstein [35] is chosen here as it follows the trends of III-V semiconductors [41]. A direct comparison of the values of $B$ taken here and by Cardona et al. [48] is not possible, since in the 16 (or 14) band model $\gamma$ is no longer directly proportional to $B$, but rather the momentum matrix element $Q$ which couples the $\Gamma_8^v, \Gamma_7^v$ and $\Gamma_8^c, \Gamma_7^c$ bands [8,48].

The results for $\beta$ as a function of $W$ exhibit a weak quadratic behaviour, decreasing with increasing well width indicated by the solid line in Fig. 3(a). This behaviour can be qualitatively accounted for by comparison to the $1/W^2$ dependence in the high barrier limit for $\beta$ given by Eppenga et al. [1] as $\beta \approx \gamma_{bulk}(\pi/W)^2$, where $\pi/W$ is the quantised wave number $k_z$ (for the 1$^{st}$ subband) in an infinite well. Although instructive as a qualitative comparison, we find that this approximation overestimates the magnitude of $\beta$ by a factor of two compared with evaluation of Eq. 8 even for the widest well (see Fig. 3(a)). This is due to the finite barriers used here and the elevated value of $\gamma$ in the bulk used in the approximation. In Fig. 3 we also compare our results for $\beta$ from Eq. 8 with the calculation of $\beta$ using $\beta = \gamma <k_z^2>$ following the approaches of Schliemann et al., Kainz et al. and Averkiev et al. [27,30,31]. Following this method we find an under estimation by a factor of ~1.2 in the widest well. We attribute the observed discrepancy to the additional two terms in $\beta$ that result from the differentiation of $\hat{\gamma}(z,\lambda)$ over the interfaces (see Eq. 8), which are comparable to a field term and an interface term similar to those appearing in $\alpha$ (Eq. 10). The overall expression for $\beta$ consisting of three terms is

$$\beta = \frac{-2P_0 B}{3}\left[\left\langle \Psi(z)\left|\frac{\partial \phi(z)}{\partial z}D_i\right|\frac{\partial \Psi(z)}{\partial z}\right\rangle + V_{UB}D_0\Psi(0)\frac{\partial \Psi(0)}{\partial z} - V_{LB}D_W\Psi(W)\frac{\partial \Psi(W)}{\partial z}\right]$$

$$\frac{-2P_0 B}{3}\left[\Psi(0)\frac{\partial \Psi(0)}{\partial z}(C_0 - V_{UB}D_0) - \Psi(W)\frac{\partial \Psi(W)}{\partial z}(C_W - V_{LB}D_W)\right]$$

$$+ \frac{2P_0 B}{3}\left[\left\langle \Psi(z)\left|\tilde{\gamma}(z,\lambda)\frac{\partial^2}{\partial z^2}\right|\Psi(z)\right\rangle\right], \quad (11)$$

where $\Psi(0)$, $\Psi(W)$, $\partial\Psi(0)/\partial z$ and $\partial\Psi(W)/\partial z$ are the values of the wave function and derivative of the wave function at the interfaces. The third term in Eq. 11 is formally equivalent to $\gamma <k_z^2>$. From inspection of the results, we find that the interface term dominates over the field term, as with $\alpha$. However, the third $\gamma <k_z^2>$ term is dominant overall contributing to ~75% of the total, justifying the approximation used by previous authors [27,28,31]. The discrepancy seen in Fig. 3 can be more specifically attributed to interface contributions, from which it is apparent that the $k$-linear BIA term has contributions from both BIA and SIA [1,41,46].

It is worth mentioning that the underlying barrier asymmetry in these structures i.e. $dV_{UB} - dV_{LB} > 0$, counteracts the direction of the field in the QW, reducing the asymmetry of the probability densities at the interfaces $\Delta\Psi_I^2$. Intuitively, one would therefore expect the Rashba coefficient to be greater in symmetric structures where the upper barrier conduction band offset $dV_{UB}$ is reduced (for doping located above the well). In varying the upper barrier alloy composition $x$, we find that although $\Delta\Psi_I^2$ does indeed increase as the underlying barrier asymmetry is reduced, the coefficients within the dominant interface term associated with the top interface, $C_0 - dV_{UB}D_0$ (see Eq. 6), which depend on band parameters, decrease in such a way as to counteract the expected behaviour. Consequently, for a fixed carrier density the Rashba coefficient is found to increase approximately linearly with asymmetry as show in Fig. 3(b) for the 20nm QW where we have reduced the upper barrier composition from $In_{0.8}Al_{0.2}Sb$ to $In_{0.85}Al_{0.15}Sb$ to match the lower barrier using interpolated band parameters from Ref. 37. The same behaviour is seen for different well widths. In contrast to the SIA coupling, we find that $\gamma$ shows negligible variation with barrier asymmetry since the contribution to $\gamma$ from the QW dominates over that from the barriers. However, similar to the result for $\alpha$, $\beta$ is also found to increase linearly with barrier asymmetry which, along with the observation of constant $\gamma$, is attributed to an increase in $<k_z^2>$. These results highlight the importance of heterostucture design for maximising or minimising $\alpha$ and $\beta$. For maximum $\alpha$, $n_s$ should be maximised for a given $W$ (consistent with the constraint of single subband occupation) with consideration given also to the barrier compositions.

## VI. SPIN-SPLITTING CALCULATIONS

The total spin-splitting $\Delta E_{Tot}[\mathbf{k}_F(\theta)]$ at the Fermi energy depends on the direction of in plane momentum $\theta(k_\parallel)$ and is calculated using the expression

$$\Delta E_{Tot}[\mathbf{k}_F(\theta)] = 2\left|<\Psi(z)|\hat{K}_{SIA} + \hat{K}_{BIA}|\Psi(z)>\right|, \quad (12)$$

where the Fermi wave vector is a function of $\theta$ given by $\mathbf{k}_F(\theta) = k_F(\cos\theta,\sin\theta,0)$ and the Fermi wave vectors are calculated from carrier densities $n_s$ obtained from the SPM solutions through the relationship $k_F = \sqrt{2\pi n_s}$. Fig. 4(a) shows the variation of the spin-splitting in the $(k_x, k_y)$ plane with $\mathbf{k}_F(\theta)$ for three 15nm QW structures of varying $k_F$ using parameters obtained in Sec. III. The contributions from BIA include both linear and cubic terms in $k$ as seen in Eq. 6. The oscillatory behaviour in $\Delta E_{Tot}[\mathbf{k}_F(\theta)]$ demonstrates that the cubic terms in $k$ from BIA contribute significantly to the total spin splitting.

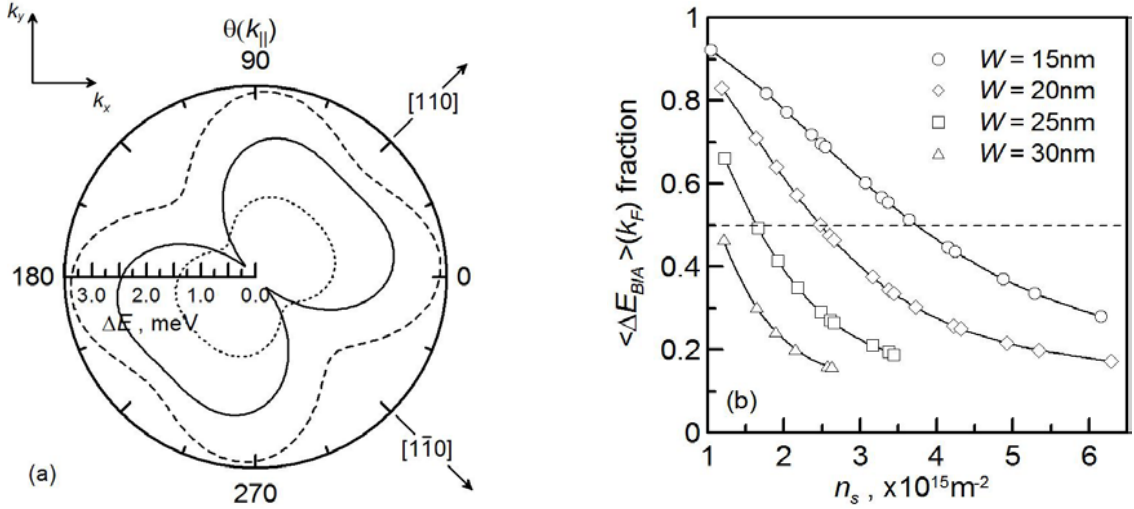

FIG. 4. (a) Calculated spin-orbit splitting at the Fermi energy as a function of $k_\parallel = (k_x, k_y)$ for three 15nm QW heterostructures with $k_F = 0.81\times10^8\text{m}^{-1}$ (dotted line), $k_F = 1.26\times10^8\text{m}^{-1}$ (solid line) and $k_F = 1.81\times10^8\text{m}^{-1}$ (dashed line) showing strong anisotropy in the [110] and [1$\bar{1}$0] directions due to the presence of both SIA and BIA. $2\beta/\gamma k_F^2 > 1$ for the data shown. (b) Fraction of the total averaged spin-splitting contributed from BIA as a function of carrier density for the different well widths. The dashed line indicates the point when both SIA and BIA contribute equally to $<\Delta E_{Tot}>(k_F)$.

The spin-splitting $\Delta E[\mathbf{k}_F(\theta)]$ from SIA and BIA separately exhibit an isotropic and a symmetric four-lobed [1] behaviour respectively with respect to $k_\parallel$. The strong anisotropy about the [110] and [1$\bar{1}$0] directions observed in Fig. 4(a) is thus due to the presence of both SO couplings simultaneously which are not additive within Eq. 12. The minimum spin-splitting is observed for $\mathbf{k}_F(\theta)$ parallel to the [1$\bar{1}$0] direction which also shows a strong non-linear dependence on $k_F$ (this is also the direction of the minimum for $\Omega[\mathbf{k}_F(\theta)]$ since $\Delta E_{Tot}[\mathbf{k}_F(\theta)] \equiv 2\hbar|\Omega[\mathbf{k}_F(\theta)]|$). The average spin-splitting is found by integrating over the Fermi circle $\mathbf{k}_F(\theta)$ giving

$$<\Delta E_{Tot}>(k_F) = 2k_F\left\{\alpha^2 + \beta^2 + \frac{k_F^4}{8}\gamma^2 - \frac{k_F^2}{2}\beta\gamma\right\}^{1/2}, \qquad (13)$$

from which the separate contributions to the spin splitting from SIA and BIA can be calculated. It can be seen from Eq. 13 that when SIA dominates, the usual spin splitting associated with the Rashba effect, $<\Delta E_{SIA}> = 2k_F|\alpha|$, is recovered. Using the SO parameters for $k_F = 1.26\times10^8\text{m}^{-1}$ we find that $<\Delta E_{BIA}>(k_F)$ contributes 14% to the total spin-splitting in the 30nm QW structure rising to 69% in the 15nm QW, due to the relative magnitudes of $\alpha$ and $\beta$ in these structures (see Fig. 2(a)). Fig. 4(b) shows the variation of the BIA contribution to the total spin-splitting with carrier density and well width demonstrating clearly that both of these parameters have a significant effect in this fraction. Furthermore, we find that the total (average) spin splitting is greater in the narrow well structures over the range of carrier densities considered, which for $k_F = 1.26\times10^8\text{m}^{-1}$ is calculated as $<\Delta E_{Tot}>(k_F) = 1.80$meV and 1.42meV for the 15nm and 30nm QW respectively. We therefore find, contrary to the results of Pfeffer et al. [15] for the InAlAs/InGaAs system, that in these InSb heterostructures the SIA cannot always be considered as the dominant mechanism as the BIA contribution is significant and dependent on well width and carrier density.

The effect of reducing the asymmetry on the coupling parameters α and β was discussed in Sec. III. For a 20nm QW, the effect of reducing the upper barrier alloy composition from $In_{0.8}Al_{0.2}Sb$ to $In_{0.85}Al_{0.15}Sb$ is manifested in a 14% reduction in $<\Delta E_{Tot}>(k_F)$ due to the reduced magnitude of coupling parameters α and β as shown in Fig. 3(b).

We note that the characteristic anisotropy of the BIA spin splitting $\Delta E_{BIA}[\mathbf{k}_F(\theta)]$ [1] exhibits a marked dependence on the ratio of β/γ and $k_F$. For particular values of β/γ and $k_F$ the anisotropy, defined as $\Delta E_{[100]}/\Delta E_{[110]}$, undergoes a π/4 phase shift compared to the earlier work of Eppenga et al. [1] for [001] grown GaAs/AlAs QWs such that the maximum spin splitting is found along the [110] direction and not the [100] direction. This macroscopic change in anisotropy occurs smoothly, and from analysis of Eq. 12 with α = 0, it can be shown that the transition in anisotropy occurs when the condition $4\beta/\gamma k_F^2 = 1$ is satisfied, at which point $\Delta E_{BIA}[\mathbf{k}_F(\theta)]$ is completely isotropic. When both SIA and BIA are present the total spin-splitting will not become isotropic due to the crossed terms involving αβ, αγ and βγ which appear in Eq. 12. The behaviour of $\Delta E_{Tot}[\mathbf{k}_F(\theta)]$ becomes completely symmetric about the [100] direction at the condition $2\beta/\gamma k_F^2 = 1$, where we find for $2\beta/\gamma k_F^2 < 1$ the minimum spin-splitting is found along the [1$\bar{1}$0] direction i.e. a π/2 phase shift compared to the data presented in Fig. 4(a) where $2\beta/\gamma k_F^2 > 1$. An optical experiment whereby the spin-splitting is measured from the side emission in the directions [110] and [1$\bar{1}$0] in a gated LED device could in principle detect such a change in symmetry from the photoluminescence or electroluminescence spectra when the carrier density is varied such that the condition of $\beta/\gamma = k_F^2/2$ is traversed. With knowledge of the carrier density, such an experiment would yield direct measurement of the ratio β/γ, which could be compared to the calculations

## V. SPIN LIFETIME CALCULATIONS

The original proposals for the SL-FET consider spin relaxation only from the k-linear SO terms α and β under the approximation of low carrier density and a negligible k-cubic contribution (γ) to the Dresselhaus splitting [28]. *Kainz et al.* [31] extended the calculations of Refs. 26, 27 and 28 in the AlGaAs/GaAs and GaAlSb/InAs material systems to more realistic carrier densities. When taking into account higher order k terms involving γ in the spin lifetime calculation *Kainz et al.* [31] observed a carrier density and well width dependence on the spin lifetime, previously overlooked [26,27]. In the following we take these issues into account.

In the $In_{1-x}Al_xSb/InSb$ systems considered SO parameters α and β are of opposite sign (shown in Fig. 2(a)). Following the theoretical results of Averkiev et al. [30] for the DP mechanism, a maximum in the spin lifetime will occur for spins injected parallel to [1$\bar{1}$0]. For a degenerate 2DEG the spin relaxation rate is given by [30]

$$\frac{1}{\tau_{s[1\bar{1}0]}} = \frac{2\tau_1}{\hbar^2}\left[(\alpha+\beta)^2 k_F^2 - \frac{1}{2}\gamma(\alpha+\beta)k_F^4 + \frac{1+\tau_3/\tau_1}{16}\gamma^2 k_F^6\right]. \quad (14)$$

$\tau_1$ and $\tau_3$ are the lifetimes associated with the harmonics of the scattering cross-section given by $\tau_n^{-1} = 1/\hbar^2 \int |U_{\mathbf{kk'}}|^2 (1-\cos n\theta).d\theta$ where $|U_{\mathbf{kk'}}|^2$ is the probability of scattering from state **k** to **k**' through angle an θ [30]. $\tau_1$ coincides with the momentum scattering lifetime $\tau_p$ related to the mobility μ of the 2DEG by $\mu = e\tau_p/m^*$ which can be obtained from simple Hall measurements. The lifetime $\tau_3$ is a more ambiguous quantity. The approximation of $\tau_3 \sim \tau_1$ is made in Ref. 31. A recent study by *Orr et al.* [38] of electron transport in modulation doped InSb/InAlSb QW heterostructures, similar or identical to those described here, demonstrated that for temperatures below 30 K, carrier mobilities are limited by long-range (small angle) remote ionized impurity scattering (RIIS). By manipulating the expression for the scattering rate associated with RIIS $\tau^{-1}_{RIIS}$ [38] to match the form of $\tau^{-1}_n$ (for n = 1), the appropriate probability of a scattering event $|U_{\mathbf{kk'}}|^2$ can be extracted from $\tau^{-1}_{RIIS}$ and substituted into the expressions for $\tau^{-1}_n$ yielding an estimate of the ratio $\tau_3/\tau_1$. Momentum scattering from ionized impurities exhibits an inherent dependence on the separation of the ionized atoms from the 2DEG (S) and $k_F$ through the Coulomb-like interaction and screening effect of mobile carriers. As a result, we find that $\tau_3/\tau_1$ varies typically from 0.12 to 0.28 as S is reduced from 20nm to 5nm respectively over the range of $k_F$ values considered. For ease of computation, in the following calculations of $\tau_{s[1\bar{1}0]}$ we use an average value of $\tau_3/\tau_1 = 0.2$).

From Eq. 14, it can be seen that for small carrier densities ($n_s < 0.5 \times 10^{15}$ m$^2$) and/or small $\gamma$ such as in wider gap materials e.g. GaAs, the $k_F^4$ and $k_F^6$ terms become negligible and $\tau_{s[1\bar{1}0]}$ is dominated by the $k_F^2$ terms involving only $\alpha$ and $\beta$ parameters. This allows $\tau_{s[1\bar{1}0]}$ to approach infinity when $\alpha = -\beta$ as demonstrated in Fig. 5 (solid line) for $\gamma = 0$ [26,27,28]. For realistic device carrier densities $n_s \sim 3 \times 10^{15}$ m$^2$ these terms are no longer negligible and should be taken into account [31].

In Fig. 2(a) we demonstrated that in the single sided δ-doped structures considered here, $\alpha$ is proportional to $n_s$. It was shown theoretically in the AlGaAs/GaAs and AlAs/InAs systems [31,51] that the spin lifetime can exhibit a strong non-monotonous behaviour with $n_s$ (seen from Eq. 14). Experimentally, the sensitivity of the spin lifetime to the carrier density, modulated either via the application of an external electric field [21,43,52] or the doping conditions [53], has been demonstrated. As a result, for the current structures it is not correct to reproduce the matching conditions for the SP-FET made by Cartoxia *et al.* [26] who assumed a constant carrier density and calculated $\tau_{s[1\bar{1}0]}$ as a function $\beta/\alpha$ with $\alpha$ as a free variable. However, modulation of $\alpha$ whilst maintaining a constant carrier density can be achieved by the use of a second 'back' gate and has been demonstrated experimentally in the work of Nitta *et al.* [7]. For this reason we consider separately the two approaches of (*i*) constant carrier density, where we assume only $\alpha$ as a free variable for a given $k_F$ in order to calculate $\tau_{s[1\bar{1}0]}$ as a function of $\beta/\alpha$ and (*ii*) varying carrier density, combining all of the results of $\alpha$, $\beta$ and $\gamma$ from the calculations in Sec. III, appropriate to the structures studied here. We emphasize that the first condition is not valid for the structures modeled but is instructive for exploring the effects of well width and carrier density on the behaviour of $\tau_{s[1\bar{1}0]}$ independently.

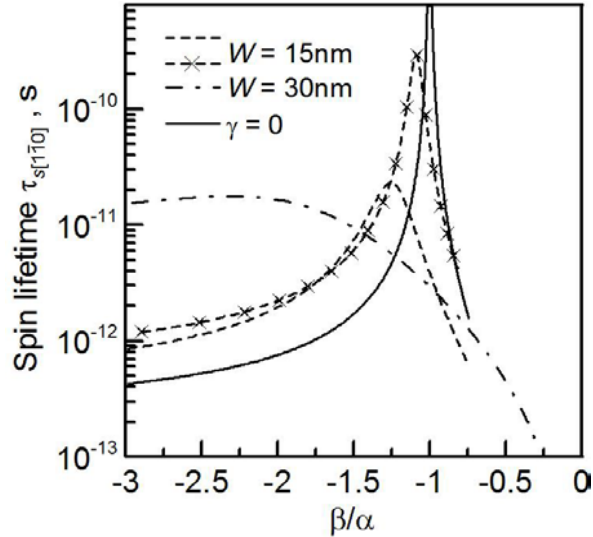

FIG. 5. Calculated spin lifetimes for the spin component directed along $[1\bar{1}0]$ as a function of the ratio of BIA and SIA k-linear parameters for a 15nm QW (dashed line) and 30nm QW (dotted-dashed line) with constant $k_F = 1.26 \times 10^8$ m$^{-1}$. Data for a 15nm QW structure with $k_F = 0.8 \times 10^8$ m$^{-1}$ (crossed and dashed line) and the result of taking $\gamma = 0$ (solid line) is plotted for comparison.

### A. Results for constant carrier density

Treating $\beta$ and $\gamma$ as fixed for a given carrier density and well width, the condition maximum spin lifetime is, from $\partial \tau_{s[1\bar{1}0]}^{-1} / \partial \alpha = 0$,

$$\alpha = \frac{1}{4}k_F^2 \gamma - \beta,  \quad (15a)$$

i.e.
$$\frac{\beta}{\alpha} = \frac{k_F^2}{k_F^2 - 4\beta/\gamma} - 1. \quad (15b)$$

It can seen that Eq. 15 reduces to the case of $\alpha = -\beta$ in the limit of small $k_F$ and $\gamma$. Fig. 5 shows the spin lifetimes $\tau_{s[1\bar{1}0]}$ calculated from Eq. 14 as a function of the ratio $\beta/\alpha$ for a 15nm (dashed line) and 30nm (dotted-dashed line) QW heterostructure using SO parameters obtained from the $\mathbf{k} \cdot \mathbf{p}$ calculations with $\tau_1 = 1.2$ps, $k_F = 1.26 \times 10^8$m$^{-1}$ and $m^* = 0.014 m_0$. SO parameters used in the calculations are $\beta = 0.0768$eVÅ and $0.0314$eVÅ, and $\gamma = 385$eVÅ$^3$ and $442$ eVÅ$^3$ for the 15nm and 30nm QW calculations respectively. From Fig. 5, it is clear that by including terms involving $\gamma$, the spin lifetime no longer approaches infinity at the conditions given in Eq. 15 but is substantially suppressed to a finite value. Substitution of Eq. 15a into Eq. 14 yields a maximum spin lifetime of

$$\tau_{s[1\bar{1}0]}^{max} = \frac{\hbar^2}{2\tau_1} \frac{16}{k_F^6 \gamma^2 (\tau_3/\tau_1)}. \quad (16)$$

This observation is consistent with a qualitative picture of the spin configuration in the 2DEG plane. In the idealized case of $\alpha = -\beta$ ($\gamma = 0$) the component along $[1\bar{1}0]$ of an injected spin is always parallel to the direction of $\mathbf{\Omega}(\mathbf{k}_\parallel)$, denoted by $\theta_{eff}[\mathbf{\Omega}(\mathbf{k}_\parallel)]$, which takes the two directions $[\bar{1}10]$ and $[1\bar{1}0]$ dependant on direction of in plane momentum $\theta(\mathbf{k}_\parallel)$. With the inclusion of $\gamma k^3$ terms in $\tau_{s[1\bar{1}0]}$, the orientation of $\mathbf{\Omega}(\mathbf{k}_\parallel)$ changes with $\theta(\mathbf{k}_\parallel)$, even when the SIA and BIA couplings are balanced according to Eq. 15. Thus the component of spin is not always aligned with $\mathbf{\Omega}(\mathbf{k}_\parallel)$ and a finite spin relaxation will result. Departure of $\theta_{eff}[\mathbf{\Omega}(\mathbf{k}_\parallel)]$ from the ideal case i.e. the directions $[\bar{1}10]$ and $[1\bar{1}0]$, increases with the magnitude of $\gamma$, consistent with Eq. 16.

The location of the maximum shows significant variation between the 15nm and 30nm QWs for the same carrier density. This is controlled by the magnitude of the ratio $\beta/\gamma$ (Eq.15), which varies by a factor of three as the well width is increased from 15nm/30nm. This shift is strongly enhanced in the InSb system due to the inherently large Dresselhaus coefficient $\gamma$. Surprisingly, the width of the peak in $\tau_{s[1\bar{1}0]}$ for the 30nm QW is significantly broadened. This broadening is attributed to the reduced magnitude of the ratio $\beta/\gamma$ in this structure which enhances the significance of the $\gamma$ terms. Perhaps more importantly from a device perspective is that because of the broadening in $\tau_{s[1\bar{1}0]}$, the potential current modulation from the 'on' to 'off' states of this structure from the tuning of $\alpha$ would be greatly reduced from that of a narrower well structure. A significant peak in $\tau_{s[1\bar{1}0]}$ for the 30nm well is recovered only for very small $n_s$, approaching the scenario of the original proposals of the SL-FET where $\gamma$ is neglected.

The effect of $k_F$, and hence $n_s$, on the variation of $\tau_{s[1\bar{1}0]}$ with $\beta/\alpha$ for the 15nm QW is indicated by the dashed and crossed-dashed lines in Fig. 5 for $k_F = 1.26$ and $0.81 \times 10^8$m$^{-1}$ respectively (each data set is generated using different values of $\beta$ and $\gamma$ according to the $\mathbf{k} \cdot \mathbf{p}$ calculations for the different $n_s$). The trend of the data agree with the approximation used by Averkiev et al. [28], whereby as $k_F$ is reduced the spin lifetime behaviour approaches that of the case of taking $\gamma = 0$ (indicated by the solid line in Fig. 5), with an increasing $\tau_{s[1\bar{1}0]}^{max}$. These results demonstrate the sensitivity of the position and magnitude of the maximum spin lifetime to $k_F$ (and $n_s$), as is evident from the $k_F^2$ and $k_F^6$ dependences in Eqs. 15 and 16 respectively.

Eq. 16 shows that $\tau_{s[1\bar{1}0]}^{max}$ is also sensitive to ratio of lifetimes $\tau_3/\tau_1$. The accuracy of $\tau_3/\tau_1$ must be kept in mind when considering the limit of Eq.16 (i.e. here we have used RIIS theory), however, it is apparent from the approach of Ref. 30 that a small ratio $\tau_3/\tau_1 \ll 1$ (which can be generated from remote doping - producing small-angle scattering) is desirable for large $\tau_{s[1\bar{1}0]}^{max}$.

**B. Results for varying carrier density**

In the structures studied here, tuning $\alpha$ with an external field also changes the carrier density. Highlighted in the Sec. III, the position of the δ-doping in the growth order with respect to the QW is now of relevance as this determines the dependence of $\alpha$ on gate bias. The following results represent the calculations of $\tau_{s[1\bar{1}0]}$ using all three SO parameters and carrier densities associated with each heterostructure. For RIIS limited transport lifetimes, $\mu$ varies explicitly with spacer thickness and doping density [39]. We address this issue by calculating the product $\tau_{s[1\bar{1}0]}\mu$ for each well width as shown in Fig. 6(a) thus removing the dependence of $\mu$ from the results. We emphasize that the results presented here are for solutions at zero bias so that the carrier density is varied through the doping conditions and not through an external field.

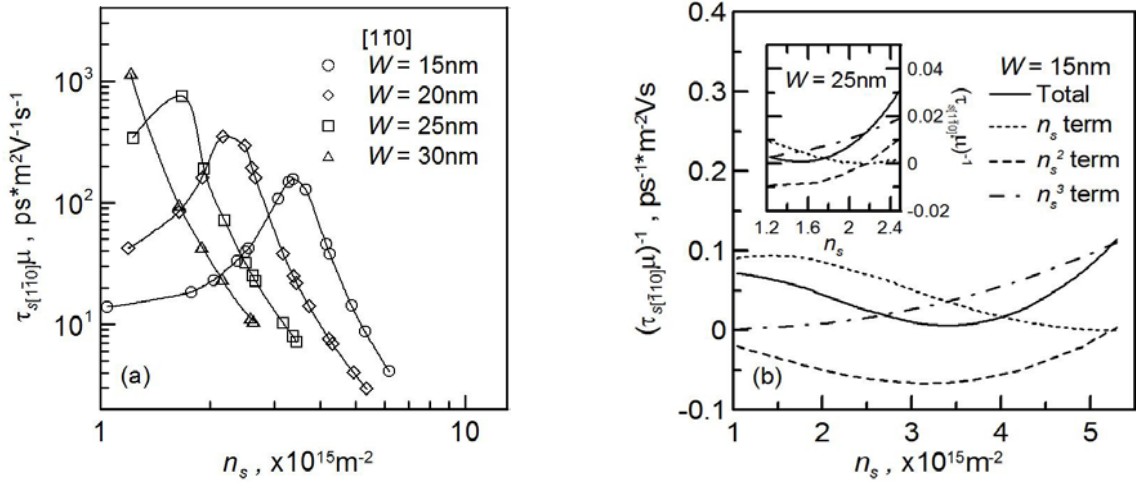

FIG. 6. (a) Results for the calculation of $\tau_{s[1\bar{1}0]}\mu$ as a function of $n_s$ for all InSb/In$_{1-x}$Al$_x$Sb structures. Data is presented on a log scale for ease of comparison. The solid lines are a guide to the eye. (b) Spin relaxation rate $(\tau_{s[1\bar{1}0]}\mu)^{-1}$ as a function of $n_s$ for the 15nm QW data showing the separate contributions from terms proportional to $n_s$, $n_s^2$ and $n_s^3$ in Eq. 13. Inset shows the contributions for the 25nm QW data for comparison where a turnover is observed at lower $n_s$ (axis are in the same units).

The results of $\tau_{s[1\bar{1}0]}\mu$ for the various well widths are presented in Fig. 6(a). Data for the 15nm, 20nm and 25nm QW structures exhibit a strong non-monotonic behaviour at moderate $n_s$ as described in Ref. 31 and 51. The lines in Fig. 6(a) represent a guide to the eye, emphasizing the turning point in the data, indicating that the SIA and BIA induced couplings have been balanced over the carrier density range considered - equivalent to the peaks seen in Fig. 5 (although the condition for maximum lifetime is now more complicated than that of Eq. 15a as $\alpha$, $\beta$ and $\gamma$ are a function of $n_s$). In Fig. 6(b) the spin relaxation rate $(\tau_{s[1\bar{1}0]}\mu)^{-1}$ is plotted against carrier density in order to elucidate the separate contributions from the terms proportional to $n_s$, $n_s^2$ and $n_s^3$ in Eq. 14. It can be seen that the turning point observed in $\tau_{s[1\bar{1}0]}\mu$ is reflected as a minimum in $(\tau_{s[1\bar{1}0]}\mu)^{-1}$ resulting from the competition between the three terms. The $n_s^2$ term has negative polarity over the range of carrier density where $\beta > \alpha$. The magnitudes of $\alpha$ and $\beta$ are comparable in the 15nm, 20nm and 25nm QW structures at moderate $n_s$ facilitating the observed peak in the spin lifetime, although notably, a maximum spin lifetime is observed at lower $n_s$ than the point at which $\alpha = -\beta$ (see Fig. 2(a)) due to the influence of $\gamma$. Accordingly, from the above observation it can be inferred that a turnover in $\tau_{s[1\bar{1}0]}\mu$ for the widest well is achieve for smaller $n_s$ than is considered here. The maximum spin lifetime increases with well width, consistent with the results of Kainz et al. [31]. This can be understood to a certain extent by comparison with the $1/n_s^3\gamma^2$ form of Eq. 16, whereby the reduced $n_s$ necessary for the suppression of $\alpha$ and subsequent appearance of a turnover, dominates over the comparatively small increase in $\gamma$ with well width.

Our results for $\tau_{s[1\bar{1}0]}\mu$ in the 20nm QW can be quantitatively compared to those of Kainz et al. [31] for a 20nm InAs/GaAlSb and GaAs/AlGaAs QWs using a momentum scattering time of $\tau_p = 0.1$ ps stated in Ref. 31 and $m^* =$

$0.014m_0$ for InSb corresponding to $\mu = 1.25 m^2V^{-1}s^{-1}$. The carrier densities at which the peak spin lifetime is observed are consistent, however, we find large disparity between the maximum values of spin lifetime estimated as ~33ns and ~10ns for the InAs and GaAs QW systems respectively from Ref. 31 compared to ~0.44ns calculated here. This is attributed to the larger Dresselhaus parameter $\gamma$ present in these InSb/InAlSb heterstructures which significantly reduces the spin lifetime. Although it is unclear which values of $\gamma$ were used in the calculations of Ref. 31, typical values of $\gamma$ in the bulk InAs and GaAs systems are $103 eV\text{Å}^3$ and $25 eV\text{Å}^3$ [20,48] respectively which would be further reduced in a heterostructure and which are substantially smaller than those calculated here.

According to the DP mechanism, the spin relaxation rate is governed by the precession of spins about an effective Larmor precession vector, during the time between momentum scattering events by $\tau_s^{-1} \propto \tau_X <\Omega^2>$ [54,55], where $\tau_X$ is the microscopic relaxation time controlling spin decoherence and $<\Omega^2>$ is the squared precession frequency averaged over the momentum distribution. The relaxation time $\tau_X$ can have contributions from mechanisms other than momentum scattering [17,55]. A meaningful cross-check for the results obtained using Eq. 14 [30] is made following the approach of Ref. 31 and assuming that the spin relaxation is effected only by those components of the precession vector which are perpendicular to the orientation of the spin.

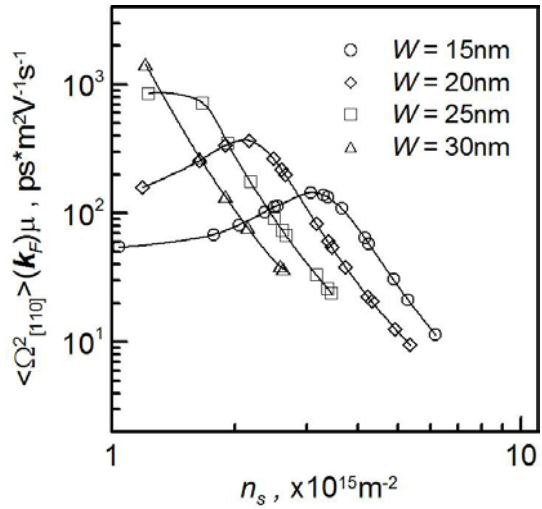

FIG. 7. Squared effective magnetic field perpendicular to spins oriented in the $[1\bar{1}0]$ direction averaged over the Fermi circle $<\Omega^2_{[110]}>(\mathbf{k}_F)$ as a function of $n_s$ for all structures of varying well widths. Data is presented on a log scale for comparison to Fig. 6(a). The solid lines are a guide to the eye.

For $[1\bar{1}0]$ oriented spins, this is the [110] direction (the component along [001] is assumed to vanish). For a degenerate 2DEG the spin relaxation rate can be expressed as [31,54]

$$\tau^{-1}_{s[1\bar{1}0]} \propto \tau_X <\Omega^2_{[110]}>(k_F) \qquad (17)$$

where $<\Omega^2_{[110]}>(k_F)$ represents $\Omega^2_{[110]}[\mathbf{k}_F(\theta)]$ averaged over the Fermi circle given by [31]

$$<\Omega^2_{[110]}>(k_F) = \frac{1}{2\pi} \int_0^{2\pi} \Omega^2_{[110]}[\mathbf{k}_F(\theta)] d\theta \qquad (18)$$

where $\mathbf{k}_F(\theta)$ is as defined in Sec. VI and $\Omega_{[110]}[\mathbf{k}_F(\theta)] = \mathbf{e} \cdot \mathbf{\Omega}_{[110]}[\mathbf{k}_F(\theta)]$ where $\mathbf{e}$ is the unit vector directed along [110]. Fig. 7 shows the results of the calculation of Eq. 17 taking $\mathbf{\Omega}(\mathbf{k}_\parallel)$ as defined in Eq. 2 and $\tau_X = \tau_p$, so as too remove the mobility dependence as done before in Fig. 6(a). An increase in broadening of the peaks is apparent compared to the data of Fig. 6(a) to the extent that the data $W = 25$nm no longer exhibits a turnover, however the location of the peaks in the

$W$ = 20 and $W$ = 15nm data show good correlation as do the magnitudes of the data in general. Evaluation of Eq. 17 yields an analytical expression which is in fact identical to Eq. 14 upon the substitution $\tau_1 = \tau_3 = \tau_p/4$. The origin of the discrepancy between the two calculations can therefore be attributed to the pre-factor of the $k_F^6$ term, which differs by 1/20 under the approximation for the ratio $\tau_3/\tau_1 = 0.2$ used in the calculations of Fig. 6. Kainz et al. [31] use the approximation $\tau_1 = \tau_3$ and thus accounts for the identical behaviour observed in $\tau_{s[1\bar{1}0]}^{-1}$ calculated from the two different methods. These results demonstrate that evaluation of Eq. 17 correctly reproduces the behaviour of $\tau_{s[1\bar{1}0]}\mu$ observed in Fig. 6(a) and indicates the validity in the approach taken by Averkiev et al. [30] and interpretations of the DP lifetime given by Glazov et al. [54] and Kainz et al. [31].

Assuming the validity of the spin lifetime expressions used in Eq. 13 and Eq. 17, the results presented here give strong indication that narrow well structures have a longer spin lifetime over the majority of realistic carrier densities ($n_s > 2 \times 10^{15}$ m$^{-2}$) compared to that of wider well structures which are greater at smaller $n_s$. Although the finer structure of the spin lifetime characteristics is dependent on all three SO parameters and the resulting effective magnetic field, we attribute this trend of spin lifetime on well width primarily to the smaller Dresselhaus parameter $\gamma$ present in the narrow wells. This parameter is found to have a significant influence on the spin lifetime maximum for both regimes of fixed and varying $n_s$.

In agreement with the results of Kaniz et al. [31] we find that spins injected parallel to the [001] and [110] directions experience approximately equal spin relaxation rates which are, typically over an order of magnitude greater than those in the $[1\bar{1}0]$ direction. This may have significant implications for spintronic devices as many spin lifetime measurements are performed by creating spins in the [001] direction through the pumping of circularly polarized light perpendicular to the surface. As a result the spin lifetime for carriers injected through a ferromagnet contact magnetized in the $[1\bar{1}0]$ direction in a Datta-Das type two terminal device may experience far less spin relaxation than is measured optically. This also highlights the importance of the orientation of the ferromagnetic contacts in such a device to achieve maximum spin lifetime.

The results of the spin lifetime for spins parallel to the [001] direction $\tau_{s[001]}\mu$ versus $n_s$ show very little variation with well widths over the range of $n_s$ considered. We find that the total contribution to the spin lifetime in this direction is controlled predominantly by the term proportional to $n_s$ in the equivalent expression for $\tau_{s[001]}$. This result is consistent with the conjecture that the well width dependence is largely due to the variation of $\gamma$ which appears only in the higher order $n_s$ terms, found to be negligible in this direction.

## VI. SUMMARY

Using an established 8 band $\mathbf{k} \cdot \mathbf{p}$ formalism [15] we have calculated the SO coupling parameters for the conduction band in $n$-InSb/In$_{1-x}$Al$_x$Sb QWs for a range of well widths and doping conditions. In contrast to previous studies [4,41], we find that in the realistic heterostructure designs considered here, for a given $n_s$ the Rashba parameter $\alpha$ is largest in wide well structures. We attribute this result to the reduced barrier alloy composition compared to that of the narrow wells. This leads to greater penetration of the wavefunction into the barriers and asymmetry of the probability densities at the interfaces $\Delta \Psi_I^2$, albeit with the caveat of a reduced range of $n_s$ over which only the first subband is occupied. Thus we find the largest Rashba coefficients can be achieved at higher carrier densities in narrow well (15nm) structures. We have demonstrated that calculation of the BIA parameter $\beta$ from the infinite well approximation can significantly overestimate the magnitude of $\beta$ whereas the more commonly employed approximation of $\beta = \gamma <k_z^2>$ accounts for only ~75% of the total magnitude since the interface contributions are neglected. We show that the BIA contribution to the spin splitting is significant and in narrow wells can exceed that of the SIA contribution which is strongly dependent on $W$ and $n_s$. We predict that under certain conditions $\Delta E_{Tot}[\mathbf{k}_F(\theta)]$ can exhibit a macroscopic $\pi/2$ change in direction of momentum giving minimum spin splitting. Experiments to verify this are suggested which would yield quantitative information of the spin splitting. Spin lifetimes have been calculated for components of spin parallel to the $[1\bar{1}0]$ direction following the work of Averkiev et al. [30] finding good qualitative agreement with the results of Kainz et al. [31] for the InAs and GaAs QW systems. We successfully reproduce the results for $\tau_{s[1\bar{1}0]}\mu$ from the precession of spins about the effective magnetic field inferred from the $\mathbf{k} \cdot \mathbf{p}$ theory [15], highlighting a discrepancy in the $k_F^6$ term from the theory of Averkiev et al. [30]. While we show that over the range of typical realistic carrier densities studied here, $\tau_{s[1\bar{1}0]}\mu$ is largest in the narrower QW, for low carrier densities large spin lifetimes are achieved in wider well structures. We have demonstrated that the inherently large SO coupling in the InSb/InAlSb system compared to other materials increases the significance of the higher order terms in the expressions for spin lifetime which as a result has considerable effect on the operating conditions of the SL-FET and the maximum achievable spin lifetime. The results of the calculations presented give important insight into the engineering of heterostructures for spintronic applications.

## ACKNOWLEDGEMENTS

The authors would like to thank C. R. Pidgeon and L. E. Golub for helpful discussions. This work was financially supported by the UK EPSRC and MOD.

---